\documentclass[secnumarabic,amssymb, nobibnotes, prl,twocolumn]{revtex4-1}
\usepackage{braket}
\usepackage{amsmath}
\usepackage{graphics}
\usepackage{graphicx}

\begin{document}

\title{Single-photon embedded eigenstates in coupled cavity-atom systems}%

\author{Michele Cotrufo}%
\email[Electronic address: ]{michele.cotrufo@utexas.edu}
\affiliation{Department of Electrical and Computer Engineering, The University of Texas at Austin, Austin, TX 78712, USA.}
\author{Andrea Al\`{u}}%
\email[Electronic address: ]{aalu@gc.cuny.edu}
\affiliation{Department of Electrical and Computer Engineering, The University of Texas at Austin, Austin, TX 78712, USA.}
\affiliation{Photonics Initiative, Advanced Science Research Center, City University of New York, New York 10031, USA.}
\affiliation{Physics Program, Graduate Center, City University of New York, New York 10016, USA.}
\affiliation{Department of Electrical Engineering, City College of The City University of New York, New York 10031, USA.}

\begin{abstract}
Confining light in open structures is a long-sought goal in nanophotonics and cavity quantum electrodynamics. Embedded eigenstates provide infinite lifetime despite the presence of available leakage channels, but in linear time-invariant systems they cannot be excited from the outside, due to reciprocity. Here, we investigate how atomic nonlinearities may support single-photon embedded eigenstates, which can be populated by a multi-photon excitation followed by internal relaxation. We calculate the system dynamics and show that photon trapping, as well as the reverse release process, can be achieved with arbitrarily high efficiencies. We also discuss the impact of loss, and a path towards the experimental verification of these concepts.
\end{abstract}

\maketitle

Trapping light in an optical cavity for times much longer than dissipative and dephasing timescales is essential to enhance light-matter interactions. Light confinement is conventionally achieved in open structures by suppressing unwanted radiation channels, e.g., with mirrors or photonic band-gap materials. However, radiation loss is never totally suppressed in conventional open resonators. Moreover, due to reciprocity, an idealized lossless cavity would also be decoupled from the outside, thus making it impossible to inject energy or to detect its internal state. Reciprocity, thus, prevents realizing linear time-invariant systems that can both efficiently collect and perfectly confine light.

Recently, there has been significant interest in embedded eigenstates (EEs), also known as bound states in the continuum \cite{Hsu2016}, optical states in open resonators that are ideally confined despite the presence of available radiation channels. Several mechanisms to form EEs have been discussed \cite{Friedrich1985,Bulgakov2008,Plotnik2011,Weimann2013,Hsu2013,Corrielli2013,Monticone2014,Mukherjee2015,Lannebere2015,Fong2017}, and their existence has been demonstrated in optical waveguides \cite{Plotnik2011,Weimann2013,Corrielli2013,Mukherjee2015}, photonic crystal slabs \cite{Hsu2013,Doeleman2018}, and nanostructures \cite{Monticone2014}. Despite their fundamentally interesting physics, in linear systems an EE is bound by reciprocity: ideal confinement implies that energy cannot be injected from the outside. Possible solutions to this problem have been proposed in classical systems by exploiting Kerr nonlinearities \cite{Lannebere2015,Bulgakov2015}. Indeed, the existence of an EE requires satisfying a strict destructive interference condition involving the geometrical parameters of the system. Thus, if the refractive index depends on the local intensity, the EE existence can be triggered dynamically, trapping part of the impinging radiation. However, due to the inherent weakness of classical nonlinearities, these approaches typically require large optical powers, and they can trap only small fractions of the incident radiation. This hinders applications of these concepts for quantum information and low-power optoelectronics, where the on-demand trapping and release of few-photons, controlled by few-photon pulses, is highly desirable.

In this Letter we propose and theoretically investigate the possibility of creating and exciting single-photon EEs by combining optical cavities and atoms. In contrast with previous approaches, here the EE excitation does not require tuning dynamically the system parameters. Indeed, thanks to atomic nonlinearities, the proposed EE inherently exists only when a single photon is present. The system is therefore “transparent” upon a single-photon excitation, but it can instead absorb a multi-photon excitation, and the subsequent internal relaxation of the system allows to populate the EE. We focus on two-photon excitation, showing that the trapping process, whereby one photon is trapped in the EE and the other is re-emitted by the system, can occur with efficiencies arbitrarily close to unity, only limited by the capability of shaping the two-photon pulse. Due to time-invariance, the inverse process is also possible: a photon stored in the EE can be released if the system is excited by a single photon. Even considering realistic losses in the system, the proposed mechanism allows storing single photons for a time much longer than the one obtainable with a single cavity, while preserving the same excitation rate. In the following, we focus on a simple system composed by one cavity and one atom. However, the approach can be readily generalized to multi-cavity systems, which may offer a more realistic framework for an experimental realization of the proposed mechanism.

\begin{figure}[htp] 	
  	\center
    \includegraphics[scale=0.58]{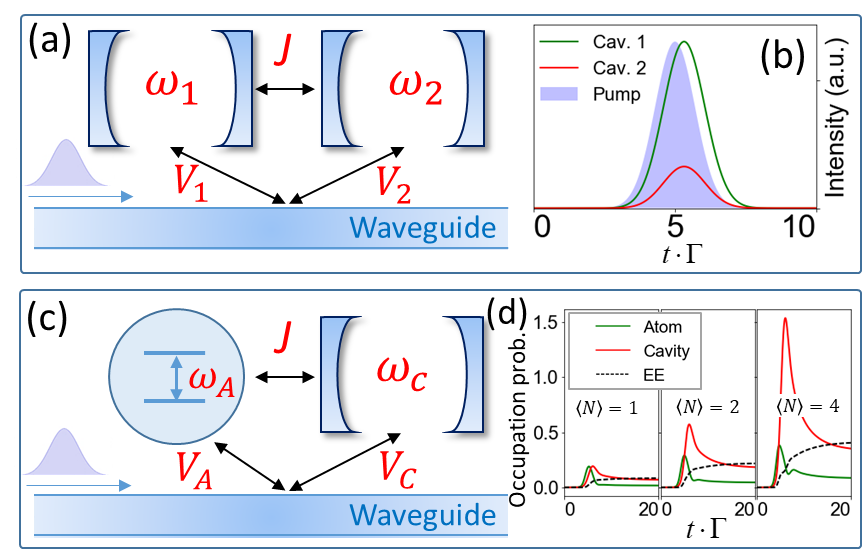}
    \caption{Embedded eigenstates due to destructive interference in classical and quantum systems. (a) Two optical cavities are mutually coupled, and they interact with a single-mode waveguide. An EE is supported when eq. \ref{eq_EE_condition} holds. (b) Time evolution of the cavity intensities (red and green lines), upon a Gaussian pulse excitation (blue shaded area). Parameters: $\omega_2=0.96\omega_1$,$V_1=\sqrt{0.01v_g\omega_1}$, $V_2=0.5V_1$. (c) Same model as in (a), but with cavity 1 replaced by a two-level system. (d) Time evolution of cavity (red line), atom (green line) and EE (black dashed line) occupation probability upon a coherent pulsed excitation containing average photon numbers $\braket{N}$ = 1, 2 and 4. System and pulse parameters are the same as in (b), with `1'$\rightarrow$`A' and `2'$\rightarrow$`C'.}\label{fig1}
\end{figure}

\textit{Model.} A straightforward condition to induce an EE in coupled systems has been proposed in \cite{Friedrich1985}. Following a similar scheme, consider the classical system (Fig. \ref{fig1}a) formed by two optical cavities resonating at frequencies $\omega_1$ and $\omega_2$, and mutually coupled with a coupling rate $J$. The cavities interact with a single-mode 1D waveguide with coupling strengths $V_1$ and $V_2$, respectively, defined such that $\Gamma_{1,2} \equiv 2 \pi V_{1,2}^2/v_g$ is the decay rate of the cavity amplitudes into the waveguide \cite{Shen2009,Roy2017}, and $v_g$ is the waveguide group velocity. We assume that, within the narrow spectral range of interest, the waveguide has a linear dispersion, and $V_1$,$V_2$ do not depend on frequency. The two-cavity system supports coupled eigenmodes that, due to the presence of the waveguide, are lossy. When the destructive interference condition
\begin{equation}
\label{eq_EE_condition}
(\omega_1-\omega_2)V_1 V_2 = J (V_1^2-V_2^2)
\end{equation}
is satisfied, one of the two coupled modes becomes lossless \cite{Friedrich1985}, realizing a bound state within the radiation continuum, while the amplitude decay rate of the other (bright) mode is  $\Gamma \equiv 2\pi (V_1^2 + V_2^2)/2 v_g$. Due to linearity, this EE exists for arbitrary intracavity intensity, i.e, for any number of photons. A light trapping mechanism based on destructive interference between cavities coupled to a waveguide was also proposed in \cite{yanik2004stopping}.

Reciprocity dictates that this EE cannot be excited from the waveguide, since a zero radiative decay rate implies that energy cannot be injected \textit{from} the outside. When a quasi-monochromatic waveguide pulse (Fig. \ref{fig1}b, blue shaded area), centered at the EE frequency, impinges on the system [set in the EE condition \ref{eq_EE_condition}], it creates transient cavity fields (green and red lines) that quickly vanish after the pulse, showing that no energy is stored. Classical nonlinearities inside one of the cavities may overcome this constraint by breaking linearity \cite{Bulgakov2015}. Here, we employ a radically different approach, assuming that one of the two resonators is inherently nonlinear at the few-photon level, as in the case of natural or artificial atoms. We thus replace one of the two cavities with a two-level atom, and the system parameters are re-labelled accordingly (Fig. \ref{fig1}c). Here we assume that the cavity and the atom couple to each other directly at a rate $J$, and they interact with the waveguide at the same point $x=0$. This formulation can be extended to the case of nonzero cavity-atom separation, where the direct coupling $J$ is replaced by the waveguide-mediated coupling, and the atom-cavity distance determines the EE condition. The existence of bound states in pairs of atoms embedded in linear waveguides has been discussed recently \cite{Facchi2016,Fang2017}, but their excitation and release dynamics have not been investigated.

The Hamiltonian of the system in Fig. \ref{fig1}c reads ($\hbar=1$)
\begin{align}
\label{eq_H_kspace}
\begin{split}
\hat{H} =  &\hat{H}_{AC} + \int_{-\infty}^{+\infty} \; dk \; \omega_k \hat{c}^\dagger_k \hat{c}_k + \\ + &\int_{-\infty}^{+\infty} \; dk \; \left[  \hat{c}^\dagger_k (V_c \hat{a} +  V_A \hat{\sigma}_-) + h.c.\right]
\end{split}
\end{align}
where $\hat{H}_{AC} = \omega_c \hat{a}^\dagger \hat{a} +\omega_A \hat{\sigma}_+  \hat{\sigma}_- + J \left(\hat{a} \hat{\sigma}_+ + h.c. \right) $ is the standard Jaynes-Cummings Hamiltonian, $\omega_c$ ($\omega_A$) is the angular frequency of the cavity (atom), $\hat{a}$ is the annihilation operator of the cavity, $\hat{\sigma}_{\pm}$ are the Pauli operators describing the two-level atom, and $\hat{c}_k$ annihilates a waveguide mode with wavevector $k$ and frequency $\omega_k$. 
Since $\hat{H}$ preserves the total number of excitations $N$, we can separately study the dynamics for each value of $N$. An $N$-excitation EE is obtained if $\hat{H}$ has an eigenstate $\ket{\Psi_{EE}^{(N)}}$ entirely localized in the atom-cavity system, i.e., $\ket{\Psi_{EE}^{(N)}} = C^{(N)} \ket{N,g} + A^{(N)} \ket{N-1,e} $, where $\ket{m,\sigma}$  denotes the state with $m$ photons in the cavity and the atom excited ($\sigma$=`\textit{e}') or in the ground ($\sigma$=`\textit{g}') state. 
Requiring that this state satisfies the time-independent Schrodinger equation leads to the condition $V_C \sqrt{N}C^{(N)}\ket{N-1,g} + V_A A^{(N)} \ket{N-1,g} + V_C \sqrt{N-1}A^{(N)}\ket{N-2,e} =0$ which, for $V_1,V_2\neq0$, can be satisfied only for $N=1$ and  $V_C C^{(N)} + V_A A^{(N)}=0$ (this condition is equivalent to (1) \cite{SuppInfo}). The atom-cavity system in Fig. \ref{fig1}c supports therefore a single-photon EE, $\ket{\Psi_{EE}^{(1)}} \propto( V_C \hat{\sigma}_+ - V_A \hat{a}^\dagger)\ket{0}$, with frequency  $\omega_{EE}^{(1)}$, and a bright single-photon state $\ket{\Psi_{B}^{(1)}} \propto( V_A \hat{\sigma}_+ + V_C \hat{a}^\dagger)\ket{0}$   with frequency  $\omega_{B}^{(1)}$ and decay rate $\Gamma \equiv 2\pi (V_A^2 + V_C^2)/2 v_g$; however, no EE exists for $N>1$, in strong contrast with the classical case.  In particular, in the two-excitation scenario $N=2$ both eigenstates are lossy but, when condition \ref{eq_EE_condition} is met, one of them with frequency $\omega_{B}^{(2)}$  has a decay rate comparable to  $\Gamma$, while the other one, at frequency $\omega_{D}^{(2)}$, has a much smaller decay rate \cite{SuppInfo}. We therefore obtain an EE with strongly nonlinear behavior on the input power: a single impinging photon cannot interact with the system (with a dynamics similar to the classical linear system, Fig. \ref{fig1}b); while a multi-photon excitation, even at the EE frequency, can excite one or more higher-energy bright states. These states then decay into single-excitation states through internal relaxation, thus trapping a portion of the excitation in the EE. The single-photon EE  $\ket{\Psi_{EE}^{(1)}}$ bears analogy to the dark states obtainable in Lambda-type atoms by finely tuning two classical control lasers \cite{Fleischhauer2005}. In our system, however, the EE existence is set only by the atom-cavity detuning and coupling strengths and, importantly, it does not require any external control laser or field to trap radiation.

We verified the expected trapping behavior by applying input-output theory \cite{Fan2010} to the Hamiltonian (\ref{eq_H_kspace}) and calculating \cite{Johansson2013} the system response upon a coherent pulsed excitation with average photon numbers $\braket{N}$  (see \cite{SuppInfo} for details). In order to compare the quantum and classical scenarios, the system and excitation parameters in Figs. \ref{fig1}(c-d) are set equal to the ones in figs. 1(a-b) (with `1'$\rightarrow$`A' and `2'$\rightarrow$`C'). Figure \ref{fig1}d shows the dynamics for three values of $\braket{N}$: after a short transient created by the pulse, the cavity and atom populations initially decay, until a steady-state with nonzero population is reached, where the energy is trapped in the EE (black dashed line). In agreement with the previous discussion, the EE features a precise balance between atom and cavity populations, following  $V_C C^{(1)} + V_A A^{(1)}=0$.

\begin{figure}[b] 	
  	\center
    \includegraphics[scale=0.39]{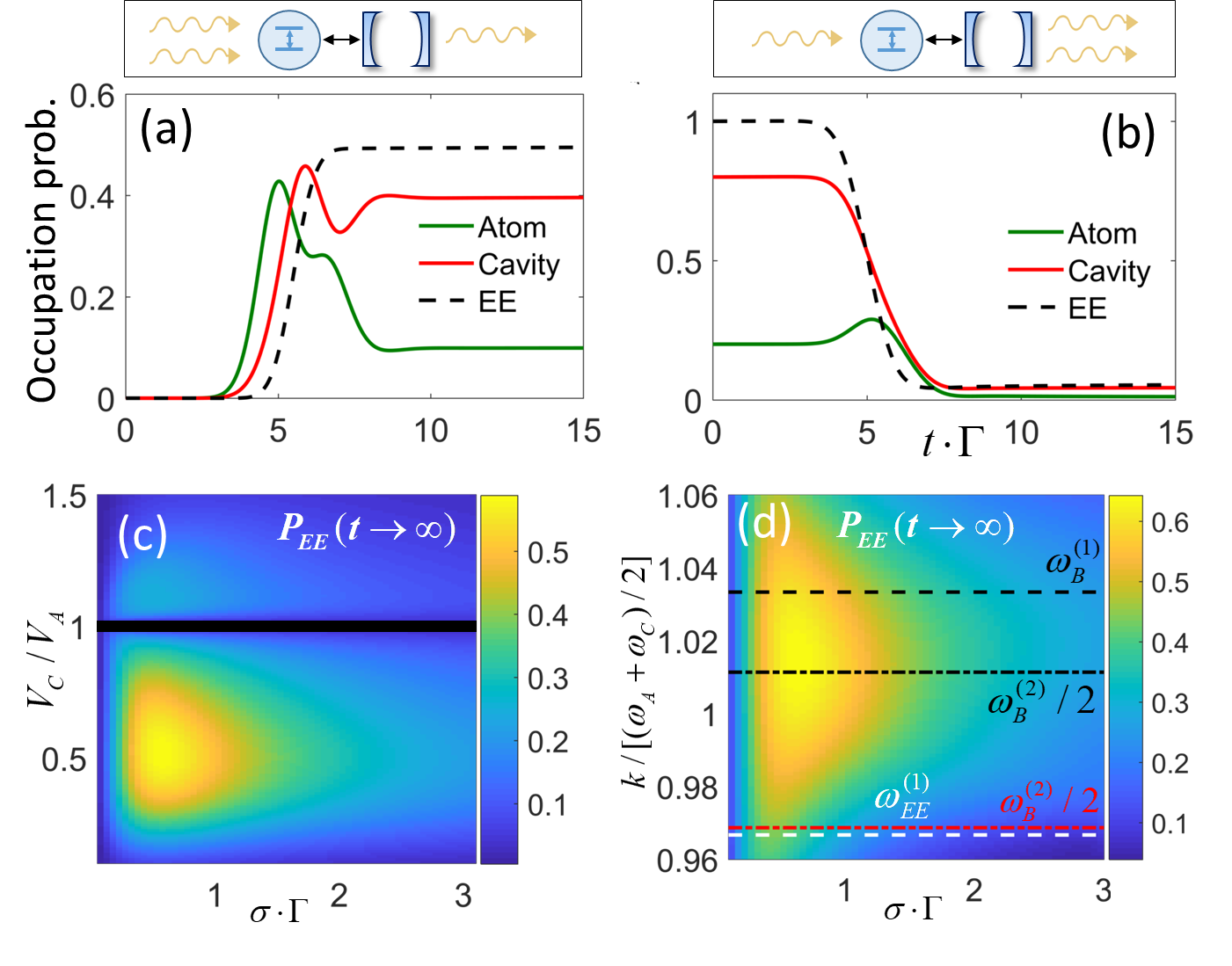}
    \caption{(a) Trapping process: a two-photon Gaussian state impinges on the system, exciting the EE with probability of about 0.5. (b) Release process: a single-photon Gaussian state impinges on an initially excited EE. System parameters in (a-b) are the same as in Fig. \ref{fig1}d, and $\sigma=1/\Gamma$, $k=\omega_B^{(1)}$ . (c-d) Excitation probability of the EE for $t\rightarrow +\infty$  versus (c) $V_C/V_A$ and $\sigma$  (for $k=\omega_B^{(1)}$) and (d) $k$ and $\sigma$  (for  $V_C/V_A=0.5$). All other parameters are the same as in (a,b). The black horizontal stripe in (c) corresponds to the values of $V_C/V_A$  that result in $J>0.1(\omega_A+\omega_c)/2$. Horizontal lines in (d) indicate the frequencies of single- and two-excitation states. }\label{fig2}
\end{figure}
Next, we consider a two-photon input, in order to verify whether trapping can occur not only for coherent excitation. Specifically, we investigate the trapping process in which two photons impinge on the system but only one emerges, as the other one is trapped in the EE.  We apply real-space formulation to the waveguide Hamiltonian  \cite{Shen2007,Shen2009} ($v_g=1$  from now on)
\begin{align}
\label{eq_Hx}
\begin{split}
&\hat{H} = \hat{H}_{AC}+\hat{H}_{WG}  +\\
&+\sqrt{2\pi}\int dx \delta(x)\left \{\left[ \hat{c}_R^\dagger(x) +\hat{c}_L^\dagger(x)\right](V_c \hat{a} +  V_A \hat{\sigma}_-) + h.c.\right\}
\end{split}
\end{align}
where $\hat{c}_R^\dagger(x)$ [$\hat{c}_L^\dagger(x)$] is the annihilation operator of a right[left]-propagating waveguide photon at position $x$, and $
\hat{H}_{WG}=-i  \int dx \;\bigg[\hat{c}^\dagger_R(x)\; \partial_x \;\hat{c}_R(x) -\hat{c}^\dagger_L(x)\; \partial_x \;\hat{c}_L(x)\bigg]$. We introduce even (`e') and odd (`o') waveguide modes \cite{Shen2009}, defined by $\hat{c}_{e/o}^\dagger(x)=\left[\hat{c}_R^\dagger(x) \pm \hat{c}_L^\dagger(-x)\right]/\sqrt{2}$, and discard the presence of the odd mode as it does not interact with the rest of the system. We focus therefore on the Hamiltonian
\begin{align}
\label{eq_Hk_e}
\begin{split}
&\hat{H}_e = \hat{H}_{AC} -i  \int dx \;\hat{c}^\dagger_e(x)\; \partial_x \;\hat{c}_e(x) +\\
&+\int \; dx \delta(x) \; \left[  \hat{c}_e^\dagger(x)(\tilde{V}_C \hat{a} +  \tilde{V}_A \hat{\sigma}_-) + h.c.\right]
\end{split}
\end{align}
where $\tilde{V}_{A/C} \equiv 2\sqrt{\pi}V_{A/C}$. By definition, an even-mode photon propagates only towards the $+x$ direction, and it corresponds to a single photon impinging on the system symmetrically from both directions. A general two-photon state of the system can be written as 
\begin{align*}
\begin{split}
\ket{\psi} = & \biggl[\int dx_1 dx_2 \chi (x_1,x_2) \hat{c}^\dagger_e(x_1)\hat{c}^\dagger_e(x_2) /\sqrt{2}+  \\ 
+ &\int dx \left(\phi_A(x) \hat{\sigma}_+ +\phi_C(x) \hat{a}^\dagger \right) \hat{c}^\dagger_e(x)+  \\ 
 + & E_{AC} \hat{a}^\dagger \hat{\sigma}_+ + E_{2C} (\hat{a}^\dagger)^2 /\sqrt{2} \biggl] \ket{0}
\end{split}
\end{align*}

 where $\chi (x_1,x_2)$  is the probability amplitude of having two photons in the waveguide at positions   $x_1$ and $x_2$,  $\phi_A(x)$ [$\phi_C(x)$] is the probability amplitude of having one excitation in the atom [cavity] and one photon in the waveguide at $x$, and   $E_{AC}$ ($E_{2C}$) is the probability amplitude of having one excitation in the cavity and one in the atom (two excitations in the cavity).  Note that  $\chi (x_1,x_2)=\chi (x_2,x_1)$ to satisfy boson statistics. By applying the time-dependent Schrodinger equation, we obtain differential equations for all probability amplitudes [18]. In the two-photon sector, the single-photon EE can be written as  $\ket{\Psi_{EE}^{(1)}} \propto( V_C \hat{\sigma}_+ - V_A \hat{a}^\dagger)\hat{c}_e^\dagger(x)\ket{0}$, describing an excited single-photon EE and a second photon at position $x$ in the waveguide. The EE occupation probability, therefore, reads $P_{EE}(t)=\int dx |V_C \phi_A(x,t)-V_A \phi_C(x,t)|^2/(V_A^2+V_C^2)$. With a home-built FDTD code \cite{Fang2017b} we calculated numerically the system dynamics upon a two-photon initial state $\ket{\psi(t=0)}=\int dx_1 dx_2 \chi_{IN}(x_1,x_2) \hat{c}^\dagger_e (x_1) \hat{c}^\dagger_e(x_2) \ket{0} / \sqrt{2}$, where $\chi_{IN}(x_1,x_2)\neq 0$ only for $x_1,x_2<0$. We initially focus on a Gaussian input defined by $\chi_{IN}(x_1,x_2) \propto \left[ f_A(x_1) f_B(x_2) + f_B(x_1)f_A(x_2) \right]$ where $f_\alpha(x) \equiv exp\left[-(x-x_\alpha)^2/2\sigma_\alpha^2\right] \cdot exp \left[i k_\alpha x \right]$. The system is set in the EE condition (same parameters as in fig. \ref{fig1} d) and the two impinging photons are in the same state ($x_A=x_B=-5/\Gamma$, $\sigma_A=\sigma_B\equiv\sigma$ and $k_A=k_B=k$) and resonant with the bright single-photon mode ($k=\omega_B^{(1)}$). At $t=5/\Gamma$ the two-photon packet reaches the atom-cavity system, exciting single- and two-excitation states of the cavity-atom system (fig. \ref{fig2}a). After a short transient, the system reaches a steady-state with a nonzero population of the atom and cavity, corresponding to  $P_{EE}\approx 0.5$. Thus, even with simple pulse shapes, the trapping process occurs with relatively high probability. As shown in Figs. \ref{fig2}(c-d), the EE excitation probability $P_{EE}(t\rightarrow \infty)$ depends sensitively on the ratio $V_C/V_A$ and on the bandwidth and carrier frequency of the two-photon pulse. In particular, for the parameters considered here, $P_{EE}(t\rightarrow \infty)$ is maximized for $V_C/V_A \approx 0.5$ (Fig. \ref{fig2}c). Note that for a fixed finite detuning $\omega_A-\omega_c\neq0$ the value of $J$ set by (1) diverges as  $V_C/V_A \rightarrow 1$, thus making the Jaynes-Cummings model not valid. We therefore removed from Fig. \ref{fig2}c the points where $J > 0.1 (\omega_c+\omega_A)/2$ (black stripes). The dependence of the EE final population on $k$ (Fig. \ref{fig2}d) confirms that the excitation of higher-energy levels happens trough the single- and two-photon bright modes of the system (dashed and dashed-dotted black lines in Fig. \ref{fig2}d). Moreover, the final EE population is maximized when $\sigma\cdot\Gamma\approx 1$ , where $\Gamma$ is the decay rate of the single-photon bright mode (which is similar to the decay rate of the two-photon bright mode \cite{SuppInfo}). This corresponds to the conjugate matching condition in classical systems, and agrees with the fact that energy can be injected in the system only through bright modes, and that a radiative state is optimally excited when the pulse duration is equal to the state lifetime \cite{Rephaeli2010}.

As the system dynamics is fully linear in a subspace with a fixed excitation number, a trapped photon can be released by simple time reversal. In this case, the initial state is 
\begin{align*}
\begin{split}
\ket{\psi(t=0)} = \int dx F(x) \hat{c}_e^\dagger(x) \ket{\Psi_{EE}^{(1)}},
\end{split}
\end{align*}
\textit{i.e.}, a photon is trapped in the EE and a single-photon wave-packet, described by $F(x) [F(x>0)=0]$, impinges on the system. Figure \ref{fig2}b shows the system evolution when $F(x)$ is a Gaussian pulse (same parameters as in Fig. \ref{fig2}a). After the single-photon pulse (not shown in the plot) reaches the system at $t=5/\Gamma$, the energy stored in the EE is released almost completely, with a residual population smaller than 0.05. 

\begin{figure}[t] 	
  	\center
    \includegraphics[scale=0.65]{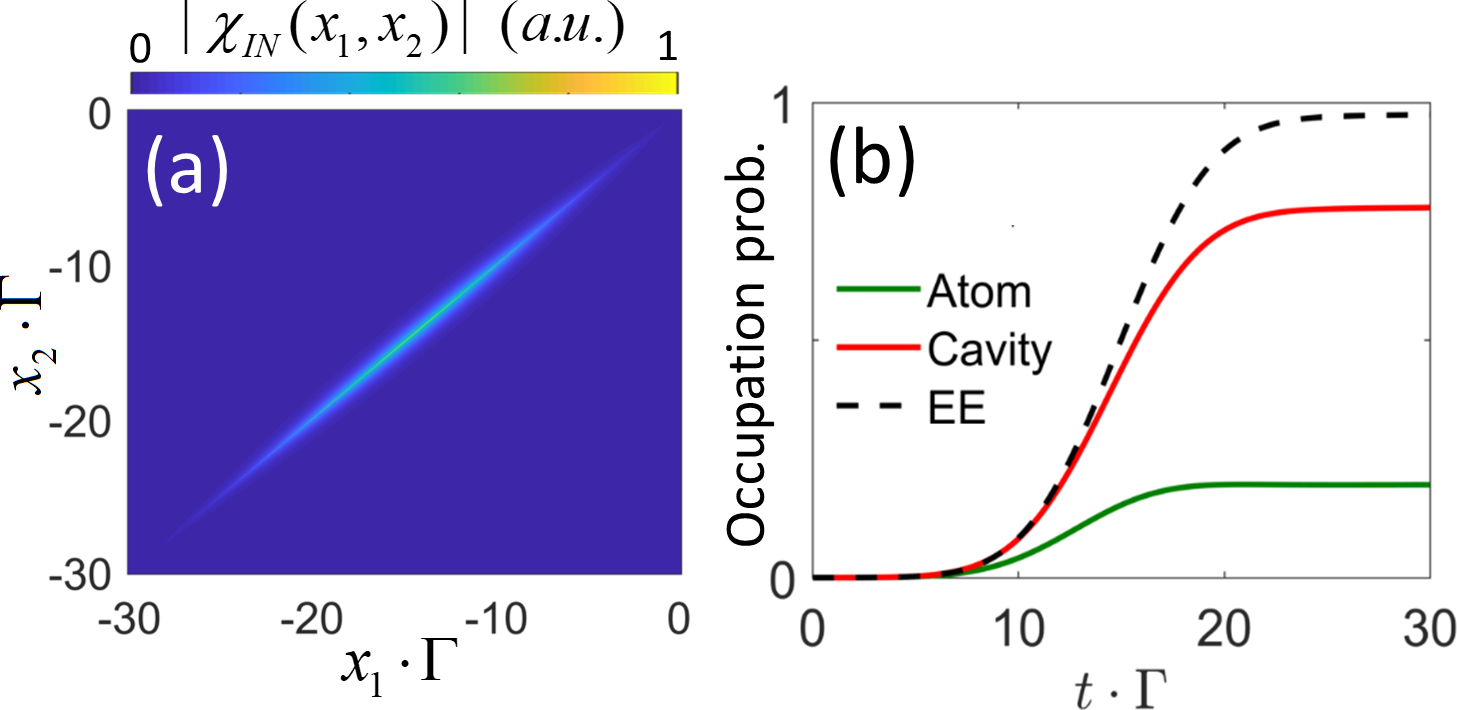}
    \caption{(a) Two-photon final state obtained when the EE is released by a single photon Gaussian packet (see text for details). (b) Evolution of the cavity-atom system when using the state in (a) as an initial two-photon state.}\label{fig3}
\end{figure}

\begin{figure*}[htp] 	
  	\center
    \includegraphics[scale=0.70]{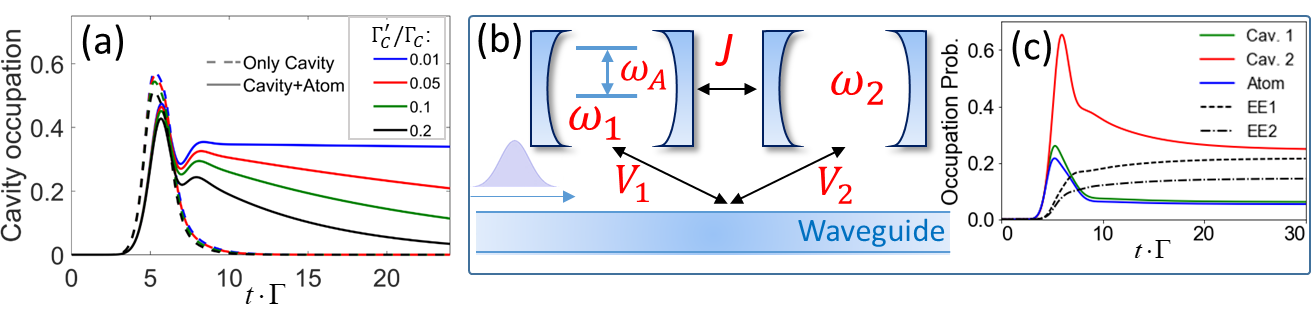}
    \caption{(a) Cavity population upon two-photon Gaussian excitation for the atom-cavity system set in the EE condition (solid lines) and for the case where only the cavity is present (dashed lines). In both cases the cavity has an additional decay rate $\Gamma_C'$. The system and excitation parameters are the same as in Fig. \ref{fig2}a. (b) A different implementation based on two coupled cavities and atom that support two single-photon EE. (c) Dynamics of the system shown in (b), set in the EE condition (5), for a coherent pulsed excitation with average photon number $\braket{N}$=2. Parameters are the same as in fig. 1b, and $J=0.003\omega_1$, $\omega_A=\omega_1$.}\label{fig4}
\end{figure*}

Much higher efficiencies in the trapping process can be obtained by using different two-photon pulse shapes. An optimal pulse can be calculated by optimizing the (inverse) release process. Indeed, if a single-photon pulse $F_{opt}(x)$  can be found such that, when impinging on an excited EE, the final state contains only two propagating photons (\textit{i.e.}, only $\chi(x_1,x_2)\neq0$ for $t\rightarrow +\infty$), the obtained two-photon state can be used as an input state to achieve unity efficiency in the trapping process. It can be shown \cite{SuppInfo} that the optimal single-photon pulse  $F_{opt}(x)$  is the solution of a  homogenous second-kind Volterra equation with a continuous kernel, which admits only the trivial solution $F_{opt}(x)=0$ [26]. However, it is possible to satisfy the requirement with arbitrarily high accuracy, albeit not identically unitary, by using Gaussian pulses with increasing spatial widths \cite{SuppInfo}. For example, when $F_{opt}(x)\propto exp[-(x-x_0)^2 + i \omega_B^{(1)}]$ with $\sigma=5/\Gamma$, after the release process the EE occupation probability (not shown here) is lower than 0.03. The corresponding final two-photon state [Fig. \ref{fig3}a, already mirrored across the origin $(x_1,x_2)=(0,0)$] is now used as an input state for the trapping process (Fig. \ref{fig3}b). The EE population now reaches values $P_{EE}(t\rightarrow \infty)\approx 0.98$, indicating a very high probability of trapping one photon. We note that, differently from the two-photon Gaussian state, the optimized $\chi_{IN}(x_1,x_2)$ (Fig. \ref{fig3}a) is squeezed along the line $x_1=x_2$; a spatial bunching of the two photons is therefore beneficial in increasing the EE excitation efficiency.

Any realistic implementation of this concept will be affected by the presence of losses, which fundamentally prevent an infinite confinement of the trapped photon.  However, as long as the decay rates of atom and cavity into the waveguide ($\Gamma_{A,C} \equiv 2\pi V^2_{A,C}/ v_g$) are much larger than these additional decay rates (denoted $\Gamma'_{A,C}$), the system allows storing a single-photon excitation in the EE for times much larger than those obtainable by a single cavity or atom. This can be clearly seen by comparing, \textit{e.g.}, the two-photon excitation of the cavity-atom system (set in the EE condition) with the case of a single cavity with same values of $\Gamma_C$ and $\Gamma'_C$ (we neglect atomic losses, as they are typically much smaller than cavity losses). As shown in fig. \ref{fig4}a, the excitation trapped in the atom-cavity system (solid lines) has now a finite lifetime that decreases as $\Gamma'_C$ increases; however, for $\Gamma'_C /\Gamma_C\leq 0.2$ the decay rate of the atom-cavity case is much smaller compared to the one of a single cavity (dashed lines). The excitation rates are instead almost the same in the two cases. We note that values of $\Gamma'_C/\Gamma_C \leq 0.1$  have been already demonstrated  experimentally in different waveguide-cavity \cite{Cai2000,Faraon2007} and waveguide-atom \cite{Arcari2014,tiecke2014nanophotonic} configurations.

So far, we have analyzed a simple system where atom and cavity are coupled to each other \textit{and} to the waveguide. This may be challenging to achieve experimentally, since, in order to have a relevant atom-cavity interaction, the atom is normally placed inside the cavity, which then makes the atom-waveguide direct interaction very weak. However, the proposed trapping mechanism can be extended to multi-cavity systems, which can be easier to implement. An example is shown in Fig. \ref{fig4}b; here, two cavities interact with each other and with the waveguide (similar to the case in Fig. \ref{fig1}a), and one atom is placed inside the first cavity, interacting with it at a rate $g$. By performing a similar analysis, we find that this system supports two different single-photon EEs, which occur when the system parameters satisfy the condition 
\begin{equation}
\label{eq_5}
(\omega_1-\omega_2)V_1 V_2 = J (V_1^2-V_2^2) + \frac{V_1^2 V_2 g^2}{V_2 J + V_1(\omega_A-\omega_2)}.
\end{equation}
Therefore, as confirmed by numerical calculations of the response of the system to a pulsed coherent excitation (Fig. \ref{fig4}c, see caption for parameters), also this structure can trap the impinging radiation in the EEs. In general, both EEs (dashed and dotted-dashed lines in Fig. \ref{fig4}c) are excited in the trapping process, but the system and pulse parameters can be tuned in order to trap radiation preferentially in one of the two.

To conclude, we proposed and theoretically investigated the dynamics of a coupled cavity-atom system supporting single-photon EEs based on atomic nonlinearities. We demonstrated that the trapping of a single photon upon a two-photon excitation can occur with arbitrarily high efficiencies if the shape of the impinging pulse can be suitably tailored. Even considering realistic losses \cite{Cai2000,Faraon2007,Arcari2014,tiecke2014nanophotonic}, the proposed mechanism allows storing single photons for a time much longer than the time needed to excite the EE, in contrast with single-cavity or single-atom configurations. Moreover, the principle is extendable to experimentally feasible systems composed of, for example, two cavities with one of them containing the atom. These findings open exciting opportunities for quantum communications and computing, as well as for attojoule optoelectronic systems.

\begin{acknowledgments}
This work has been supported by the Air Force Office of Scientific Research.
\end{acknowledgments}


\begin{thebibliography}{29}%
\makeatletter
\providecommand \@ifxundefined [1]{%
 \@ifx{#1\undefined}
}%
\providecommand \@ifnum [1]{%
 \ifnum #1\expandafter \@firstoftwo
 \else \expandafter \@secondoftwo
 \fi
}%
\providecommand \@ifx [1]{%
 \ifx #1\expandafter \@firstoftwo
 \else \expandafter \@secondoftwo
 \fi
}%
\providecommand \natexlab [1]{#1}%
\providecommand \enquote  [1]{``#1''}%
\providecommand \bibnamefont  [1]{#1}%
\providecommand \bibfnamefont [1]{#1}%
\providecommand \citenamefont [1]{#1}%
\providecommand \href@noop [0]{\@secondoftwo}%
\providecommand \href [0]{\begingroup \@sanitize@url \@href}%
\providecommand \@href[1]{\@@startlink{#1}\@@href}%
\providecommand \@@href[1]{\endgroup#1\@@endlink}%
\providecommand \@sanitize@url [0]{\catcode `\\12\catcode `\$12\catcode
  `\&12\catcode `\#12\catcode `\^12\catcode `\_12\catcode `\%12\relax}%
\providecommand \@@startlink[1]{}%
\providecommand \@@endlink[0]{}%
\providecommand \url  [0]{\begingroup\@sanitize@url \@url }%
\providecommand \@url [1]{\endgroup\@href {#1}{\urlprefix }}%
\providecommand \urlprefix  [0]{URL }%
\providecommand \Eprint [0]{\href }%
\providecommand \doibase [0]{http://dx.doi.org/}%
\providecommand \selectlanguage [0]{\@gobble}%
\providecommand \bibinfo  [0]{\@secondoftwo}%
\providecommand \bibfield  [0]{\@secondoftwo}%
\providecommand \translation [1]{[#1]}%
\providecommand \BibitemOpen [0]{}%
\providecommand \bibitemStop [0]{}%
\providecommand \bibitemNoStop [0]{.\EOS\space}%
\providecommand \EOS [0]{\spacefactor3000\relax}%
\providecommand \BibitemShut  [1]{\csname bibitem#1\endcsname}%
\let\auto@bib@innerbib\@empty
\bibitem [{\citenamefont {Hsu}\ \emph {et~al.}(2016)\citenamefont {Hsu},
  \citenamefont {Zhen}, \citenamefont {Stone}, \citenamefont {Joannopoulos},\
  and\ \citenamefont {Solja{\v{c}}i{\'c}}}]{Hsu2016}%
  \BibitemOpen
  \bibfield  {author} {\bibinfo {author} {\bibfnamefont {C.~W.}\ \bibnamefont
  {Hsu}}, \bibinfo {author} {\bibfnamefont {B.}~\bibnamefont {Zhen}}, \bibinfo
  {author} {\bibfnamefont {A.~D.}\ \bibnamefont {Stone}}, \bibinfo {author}
  {\bibfnamefont {J.~D.}\ \bibnamefont {Joannopoulos}}, \ and\ \bibinfo
  {author} {\bibfnamefont {M.}~\bibnamefont {Solja{\v{c}}i{\'c}}},\ }\href@noop
  {} {\bibfield  {journal} {\bibinfo  {journal} {Nat. Rev. Mat.}\ }\textbf
  {\bibinfo {volume} {1}},\ \bibinfo {pages} {16048} (\bibinfo {year}
  {2016})}\BibitemShut {NoStop}%
\bibitem [{\citenamefont {Friedrich}\ and\ \citenamefont
  {Wintgen}(1985)}]{Friedrich1985}%
  \BibitemOpen
  \bibfield  {author} {\bibinfo {author} {\bibfnamefont {H.}~\bibnamefont
  {Friedrich}}\ and\ \bibinfo {author} {\bibfnamefont {D.}~\bibnamefont
  {Wintgen}},\ }\href@noop {} {\bibfield  {journal} {\bibinfo  {journal} {Phys.
  Rev. A}\ }\textbf {\bibinfo {volume} {32}},\ \bibinfo {pages} {3231}
  (\bibinfo {year} {1985})}\BibitemShut {NoStop}%
\bibitem [{\citenamefont {Bulgakov}\ and\ \citenamefont
  {Sadreev}(2008)}]{Bulgakov2008}%
  \BibitemOpen
  \bibfield  {author} {\bibinfo {author} {\bibfnamefont {E.~N.}\ \bibnamefont
  {Bulgakov}}\ and\ \bibinfo {author} {\bibfnamefont {A.~F.}\ \bibnamefont
  {Sadreev}},\ }\href@noop {} {\bibfield  {journal} {\bibinfo  {journal} {Phys.
  Rev. B}\ }\textbf {\bibinfo {volume} {78}},\ \bibinfo {pages} {075105}
  (\bibinfo {year} {2008})}\BibitemShut {NoStop}%
\bibitem [{\citenamefont {Plotnik}\ \emph {et~al.}(2011)\citenamefont
  {Plotnik}, \citenamefont {Peleg}, \citenamefont {Dreisow}, \citenamefont
  {Heinrich}, \citenamefont {Nolte}, \citenamefont {Szameit},\ and\
  \citenamefont {Segev}}]{Plotnik2011}%
  \BibitemOpen
  \bibfield  {author} {\bibinfo {author} {\bibfnamefont {Y.}~\bibnamefont
  {Plotnik}}, \bibinfo {author} {\bibfnamefont {O.}~\bibnamefont {Peleg}},
  \bibinfo {author} {\bibfnamefont {F.}~\bibnamefont {Dreisow}}, \bibinfo
  {author} {\bibfnamefont {M.}~\bibnamefont {Heinrich}}, \bibinfo {author}
  {\bibfnamefont {S.}~\bibnamefont {Nolte}}, \bibinfo {author} {\bibfnamefont
  {A.}~\bibnamefont {Szameit}}, \ and\ \bibinfo {author} {\bibfnamefont
  {M.}~\bibnamefont {Segev}},\ }\href@noop {} {\bibfield  {journal} {\bibinfo
  {journal} {Phys. Rev. Lett.}\ }\textbf {\bibinfo {volume} {107}},\ \bibinfo
  {pages} {183901} (\bibinfo {year} {2011})}\BibitemShut {NoStop}%
\bibitem [{\citenamefont {Weimann}\ \emph {et~al.}(2013)\citenamefont
  {Weimann}, \citenamefont {Xu}, \citenamefont {Keil}, \citenamefont
  {Miroshnichenko}, \citenamefont {T{\"u}nnermann}, \citenamefont {Nolte},
  \citenamefont {Sukhorukov}, \citenamefont {Szameit},\ and\ \citenamefont
  {Kivshar}}]{Weimann2013}%
  \BibitemOpen
  \bibfield  {author} {\bibinfo {author} {\bibfnamefont {S.}~\bibnamefont
  {Weimann}}, \bibinfo {author} {\bibfnamefont {Y.}~\bibnamefont {Xu}},
  \bibinfo {author} {\bibfnamefont {R.}~\bibnamefont {Keil}}, \bibinfo {author}
  {\bibfnamefont {A.~E.}\ \bibnamefont {Miroshnichenko}}, \bibinfo {author}
  {\bibfnamefont {A.}~\bibnamefont {T{\"u}nnermann}}, \bibinfo {author}
  {\bibfnamefont {S.}~\bibnamefont {Nolte}}, \bibinfo {author} {\bibfnamefont
  {A.~A.}\ \bibnamefont {Sukhorukov}}, \bibinfo {author} {\bibfnamefont
  {A.}~\bibnamefont {Szameit}}, \ and\ \bibinfo {author} {\bibfnamefont
  {Y.~S.}\ \bibnamefont {Kivshar}},\ }\href@noop {} {\bibfield  {journal}
  {\bibinfo  {journal} {Phys. Rev. Lett.}\ }\textbf {\bibinfo {volume} {111}},\
  \bibinfo {pages} {240403} (\bibinfo {year} {2013})}\BibitemShut {NoStop}%
\bibitem [{\citenamefont {Hsu}\ \emph {et~al.}(2013)\citenamefont {Hsu},
  \citenamefont {Zhen}, \citenamefont {Lee}, \citenamefont {Chua},
  \citenamefont {Johnson}, \citenamefont {Joannopoulos},\ and\ \citenamefont
  {Solja{\v{c}}i{\'c}}}]{Hsu2013}%
  \BibitemOpen
  \bibfield  {author} {\bibinfo {author} {\bibfnamefont {C.~W.}\ \bibnamefont
  {Hsu}}, \bibinfo {author} {\bibfnamefont {B.}~\bibnamefont {Zhen}}, \bibinfo
  {author} {\bibfnamefont {J.}~\bibnamefont {Lee}}, \bibinfo {author}
  {\bibfnamefont {S.-L.}\ \bibnamefont {Chua}}, \bibinfo {author}
  {\bibfnamefont {S.~G.}\ \bibnamefont {Johnson}}, \bibinfo {author}
  {\bibfnamefont {J.~D.}\ \bibnamefont {Joannopoulos}}, \ and\ \bibinfo
  {author} {\bibfnamefont {M.}~\bibnamefont {Solja{\v{c}}i{\'c}}},\ }\href@noop
  {} {\bibfield  {journal} {\bibinfo  {journal} {Nature}\ }\textbf {\bibinfo
  {volume} {499}},\ \bibinfo {pages} {188} (\bibinfo {year}
  {2013})}\BibitemShut {NoStop}%
\bibitem [{\citenamefont {Corrielli}\ \emph {et~al.}(2013)\citenamefont
  {Corrielli}, \citenamefont {Della~Valle}, \citenamefont {Crespi},
  \citenamefont {Osellame},\ and\ \citenamefont {Longhi}}]{Corrielli2013}%
  \BibitemOpen
  \bibfield  {author} {\bibinfo {author} {\bibfnamefont {G.}~\bibnamefont
  {Corrielli}}, \bibinfo {author} {\bibfnamefont {G.}~\bibnamefont
  {Della~Valle}}, \bibinfo {author} {\bibfnamefont {A.}~\bibnamefont {Crespi}},
  \bibinfo {author} {\bibfnamefont {R.}~\bibnamefont {Osellame}}, \ and\
  \bibinfo {author} {\bibfnamefont {S.}~\bibnamefont {Longhi}},\ }\href@noop {}
  {\bibfield  {journal} {\bibinfo  {journal} {Phys. Rev. Lett.}\ }\textbf
  {\bibinfo {volume} {111}},\ \bibinfo {pages} {220403} (\bibinfo {year}
  {2013})}\BibitemShut {NoStop}%
\bibitem [{\citenamefont {Monticone}\ and\ \citenamefont
  {Alu}(2014)}]{Monticone2014}%
  \BibitemOpen
  \bibfield  {author} {\bibinfo {author} {\bibfnamefont {F.}~\bibnamefont
  {Monticone}}\ and\ \bibinfo {author} {\bibfnamefont {A.}~\bibnamefont
  {Alu}},\ }\href@noop {} {\bibfield  {journal} {\bibinfo  {journal} {Phys.
  Rev. Lett.}\ }\textbf {\bibinfo {volume} {112}},\ \bibinfo {pages} {213903}
  (\bibinfo {year} {2014})}\BibitemShut {NoStop}%
\bibitem [{\citenamefont {Mukherjee}\ \emph {et~al.}(2015)\citenamefont
  {Mukherjee}, \citenamefont {Spracklen}, \citenamefont {Choudhury},
  \citenamefont {Goldman}, \citenamefont {{\"O}hberg}, \citenamefont
  {Andersson},\ and\ \citenamefont {Thomson}}]{Mukherjee2015}%
  \BibitemOpen
  \bibfield  {author} {\bibinfo {author} {\bibfnamefont {S.}~\bibnamefont
  {Mukherjee}}, \bibinfo {author} {\bibfnamefont {A.}~\bibnamefont
  {Spracklen}}, \bibinfo {author} {\bibfnamefont {D.}~\bibnamefont
  {Choudhury}}, \bibinfo {author} {\bibfnamefont {N.}~\bibnamefont {Goldman}},
  \bibinfo {author} {\bibfnamefont {P.}~\bibnamefont {{\"O}hberg}}, \bibinfo
  {author} {\bibfnamefont {E.}~\bibnamefont {Andersson}}, \ and\ \bibinfo
  {author} {\bibfnamefont {R.~R.}\ \bibnamefont {Thomson}},\ }\href@noop {}
  {\bibfield  {journal} {\bibinfo  {journal} {Phys. Rev. Lett.}\ }\textbf
  {\bibinfo {volume} {114}},\ \bibinfo {pages} {245504} (\bibinfo {year}
  {2015})}\BibitemShut {NoStop}%
\bibitem [{\citenamefont {Lanneb{\`e}re}\ and\ \citenamefont
  {Silveirinha}(2015)}]{Lannebere2015}%
  \BibitemOpen
  \bibfield  {author} {\bibinfo {author} {\bibfnamefont {S.}~\bibnamefont
  {Lanneb{\`e}re}}\ and\ \bibinfo {author} {\bibfnamefont {M.~G.}\ \bibnamefont
  {Silveirinha}},\ }\href@noop {} {\bibfield  {journal} {\bibinfo  {journal}
  {Nat. Commun.}\ }\textbf {\bibinfo {volume} {6}},\ \bibinfo {pages} {8766}
  (\bibinfo {year} {2015})}\BibitemShut {NoStop}%
\bibitem [{\citenamefont {Fong}\ and\ \citenamefont {Law}(2017)}]{Fong2017}%
  \BibitemOpen
  \bibfield  {author} {\bibinfo {author} {\bibfnamefont {P.}~\bibnamefont
  {Fong}}\ and\ \bibinfo {author} {\bibfnamefont {C.}~\bibnamefont {Law}},\
  }\href@noop {} {\bibfield  {journal} {\bibinfo  {journal} {Phys. Rev. A}\
  }\textbf {\bibinfo {volume} {96}},\ \bibinfo {pages} {023842} (\bibinfo
  {year} {2017})}\BibitemShut {NoStop}%
\bibitem [{\citenamefont {{Doeleman, H.M and Monticone, F. and den Hollander,
  W. and Al{\`{u}}}}\ and\ \citenamefont {Koenderink}(2018)}]{Doeleman2018}%
  \BibitemOpen
  \bibfield  {author} {\bibinfo {author} {\bibfnamefont {A.}~\bibnamefont
  {{Doeleman, H.M and Monticone, F. and den Hollander, W. and Al{\`{u}}}}}\
  and\ \bibinfo {author} {\bibfnamefont {A.~F.}\ \bibnamefont {Koenderink}},\
  }\href@noop {} {\bibfield  {journal} {\bibinfo  {journal} {Nat. Photonics, in
  press}\ } (\bibinfo {year} {2018})}\BibitemShut {NoStop}%
\bibitem [{\citenamefont {Bulgakov}\ \emph {et~al.}(2015)\citenamefont
  {Bulgakov}, \citenamefont {Pichugin},\ and\ \citenamefont
  {Sadreev}}]{Bulgakov2015}%
  \BibitemOpen
  \bibfield  {author} {\bibinfo {author} {\bibfnamefont {E.}~\bibnamefont
  {Bulgakov}}, \bibinfo {author} {\bibfnamefont {K.}~\bibnamefont {Pichugin}},
  \ and\ \bibinfo {author} {\bibfnamefont {A.}~\bibnamefont {Sadreev}},\
  }\href@noop {} {\bibfield  {journal} {\bibinfo  {journal} {Opt. Express}\
  }\textbf {\bibinfo {volume} {23}},\ \bibinfo {pages} {22520} (\bibinfo {year}
  {2015})}\BibitemShut {NoStop}%
\bibitem [{\citenamefont {Shen}\ \emph {et~al.}(2009)\citenamefont {Shen},
  \citenamefont {Fan} \emph {et~al.}}]{Shen2009}%
  \BibitemOpen
  \bibfield  {author} {\bibinfo {author} {\bibfnamefont {J.-T.}\ \bibnamefont
  {Shen}}, \bibinfo {author} {\bibfnamefont {S.}~\bibnamefont {Fan}},  \emph
  {et~al.},\ }\href@noop {} {\bibfield  {journal} {\bibinfo  {journal} {Phys.
  Rev. A}\ }\textbf {\bibinfo {volume} {79}},\ \bibinfo {pages} {023837}
  (\bibinfo {year} {2009})}\BibitemShut {NoStop}%
\bibitem [{\citenamefont {Roy}\ \emph {et~al.}(2017)\citenamefont {Roy},
  \citenamefont {Wilson},\ and\ \citenamefont {Firstenberg}}]{Roy2017}%
  \BibitemOpen
  \bibfield  {author} {\bibinfo {author} {\bibfnamefont {D.}~\bibnamefont
  {Roy}}, \bibinfo {author} {\bibfnamefont {C.~M.}\ \bibnamefont {Wilson}}, \
  and\ \bibinfo {author} {\bibfnamefont {O.}~\bibnamefont {Firstenberg}},\
  }\href@noop {} {\bibfield  {journal} {\bibinfo  {journal} {Rev. Mod. Phys.}\
  }\textbf {\bibinfo {volume} {89}},\ \bibinfo {pages} {021001} (\bibinfo
  {year} {2017})}\BibitemShut {NoStop}%
\bibitem [{\citenamefont {Yanik}\ \emph {et~al.}(2004)\citenamefont {Yanik},
  \citenamefont {Suh}, \citenamefont {Wang},\ and\ \citenamefont
  {Fan}}]{yanik2004stopping}%
  \BibitemOpen
  \bibfield  {author} {\bibinfo {author} {\bibfnamefont {M.~F.}\ \bibnamefont
  {Yanik}}, \bibinfo {author} {\bibfnamefont {W.}~\bibnamefont {Suh}}, \bibinfo
  {author} {\bibfnamefont {Z.}~\bibnamefont {Wang}}, \ and\ \bibinfo {author}
  {\bibfnamefont {S.}~\bibnamefont {Fan}},\ }\href@noop {} {\bibfield
  {journal} {\bibinfo  {journal} {Phys. Rev. Lett.}\ }\textbf {\bibinfo
  {volume} {93}},\ \bibinfo {pages} {233903} (\bibinfo {year}
  {2004})}\BibitemShut {NoStop}%
\bibitem [{\citenamefont {Facchi}\ \emph {et~al.}(2016)\citenamefont {Facchi},
  \citenamefont {Kim}, \citenamefont {Pascazio}, \citenamefont {Pepe},
  \citenamefont {Pomarico},\ and\ \citenamefont {Tufarelli}}]{Facchi2016}%
  \BibitemOpen
  \bibfield  {author} {\bibinfo {author} {\bibfnamefont {P.}~\bibnamefont
  {Facchi}}, \bibinfo {author} {\bibfnamefont {M.}~\bibnamefont {Kim}},
  \bibinfo {author} {\bibfnamefont {S.}~\bibnamefont {Pascazio}}, \bibinfo
  {author} {\bibfnamefont {F.~V.}\ \bibnamefont {Pepe}}, \bibinfo {author}
  {\bibfnamefont {D.}~\bibnamefont {Pomarico}}, \ and\ \bibinfo {author}
  {\bibfnamefont {T.}~\bibnamefont {Tufarelli}},\ }\href@noop {} {\bibfield
  {journal} {\bibinfo  {journal} {Phys. Rev. A}\ }\textbf {\bibinfo {volume}
  {94}},\ \bibinfo {pages} {043839} (\bibinfo {year} {2016})}\BibitemShut
  {NoStop}%
\bibitem [{\citenamefont {Fang}\ \emph {et~al.}(2017)\citenamefont {Fang},
  \citenamefont {Baranger} \emph {et~al.}}]{Fang2017}%
  \BibitemOpen
  \bibfield  {author} {\bibinfo {author} {\bibfnamefont {Y.-L.~L.}\
  \bibnamefont {Fang}}, \bibinfo {author} {\bibfnamefont {H.~U.}\ \bibnamefont
  {Baranger}},  \emph {et~al.},\ }\href@noop {} {\bibfield  {journal} {\bibinfo
   {journal} {Phys. Rev. A}\ }\textbf {\bibinfo {volume} {96}},\ \bibinfo
  {pages} {013842} (\bibinfo {year} {2017})}\BibitemShut {NoStop}%
\bibitem [{Sup()}]{SuppInfo}%
  \BibitemOpen
  \href@noop {} {\bibinfo  {journal} {Supplementary Information}\ }\BibitemShut
  {NoStop}%
\bibitem [{\citenamefont {Fleischhauer}\ \emph {et~al.}(2005)\citenamefont
  {Fleischhauer}, \citenamefont {Imamoglu},\ and\ \citenamefont
  {Marangos}}]{Fleischhauer2005}%
  \BibitemOpen
\bibfield  {journal} {  }\bibfield  {author} {\bibinfo {author} {\bibfnamefont
  {M.}~\bibnamefont {Fleischhauer}}, \bibinfo {author} {\bibfnamefont
  {A.}~\bibnamefont {Imamoglu}}, \ and\ \bibinfo {author} {\bibfnamefont
  {J.~P.}\ \bibnamefont {Marangos}},\ }\href@noop {} {\bibfield  {journal}
  {\bibinfo  {journal} {Rev. Mod. Phys.}\ }\textbf {\bibinfo {volume} {77}},\
  \bibinfo {pages} {633} (\bibinfo {year} {2005})}\BibitemShut {NoStop}%
\bibitem [{\citenamefont {Fan}\ \emph {et~al.}(2010)\citenamefont {Fan},
  \citenamefont {Kocaba{\c{s}}},\ and\ \citenamefont {Shen}}]{Fan2010}%
  \BibitemOpen
  \bibfield  {author} {\bibinfo {author} {\bibfnamefont {S.}~\bibnamefont
  {Fan}}, \bibinfo {author} {\bibfnamefont {{\c{S}}.~E.}\ \bibnamefont
  {Kocaba{\c{s}}}}, \ and\ \bibinfo {author} {\bibfnamefont {J.-T.}\
  \bibnamefont {Shen}},\ }\href@noop {} {\bibfield  {journal} {\bibinfo
  {journal} {Phys. Rev. A}\ }\textbf {\bibinfo {volume} {82}},\ \bibinfo
  {pages} {063821} (\bibinfo {year} {2010})}\BibitemShut {NoStop}%
\bibitem [{\citenamefont {Johansson}\ \emph {et~al.}(2012)\citenamefont
  {Johansson}, \citenamefont {Nation},\ and\ \citenamefont
  {Nori}}]{Johansson2013}%
  \BibitemOpen
  \bibfield  {author} {\bibinfo {author} {\bibfnamefont {J.}~\bibnamefont
  {Johansson}}, \bibinfo {author} {\bibfnamefont {P.}~\bibnamefont {Nation}}, \
  and\ \bibinfo {author} {\bibfnamefont {F.}~\bibnamefont {Nori}},\ }\href@noop
  {} {\bibfield  {journal} {\bibinfo  {journal} {Comput. Phys. Commun.}\
  }\textbf {\bibinfo {volume} {183}},\ \bibinfo {pages} {1760} (\bibinfo {year}
  {2012})}\BibitemShut {NoStop}%
\bibitem [{\citenamefont {Shen}\ and\ \citenamefont {Fan}(2007)}]{Shen2007}%
  \BibitemOpen
  \bibfield  {author} {\bibinfo {author} {\bibfnamefont {J.-T.}\ \bibnamefont
  {Shen}}\ and\ \bibinfo {author} {\bibfnamefont {S.}~\bibnamefont {Fan}},\
  }\href@noop {} {\bibfield  {journal} {\bibinfo  {journal} {Phys. Rev. A}\
  }\textbf {\bibinfo {volume} {76}},\ \bibinfo {pages} {062709} (\bibinfo
  {year} {2007})}\BibitemShut {NoStop}%
\bibitem [{\citenamefont {Fang}(2017)}]{Fang2017b}%
  \BibitemOpen
  \bibfield  {author} {\bibinfo {author} {\bibfnamefont {Y.-L.~L.}\
  \bibnamefont {Fang}},\ }\href@noop {} {\bibfield  {journal} {\bibinfo
  {journal} {arXiv preprint arXiv:1707.05943}\ } (\bibinfo {year}
  {2017})}\BibitemShut {NoStop}%
\bibitem [{\citenamefont {Rephaeli}\ \emph {et~al.}(2010)\citenamefont
  {Rephaeli}, \citenamefont {Shen},\ and\ \citenamefont {Fan}}]{Rephaeli2010}%
  \BibitemOpen
  \bibfield  {author} {\bibinfo {author} {\bibfnamefont {E.}~\bibnamefont
  {Rephaeli}}, \bibinfo {author} {\bibfnamefont {J.~T.}\ \bibnamefont {Shen}},
  \ and\ \bibinfo {author} {\bibfnamefont {S.}~\bibnamefont {Fan}},\
  }\href@noop {} {\bibfield  {journal} {\bibinfo  {journal} {Phys. Rev. A}\
  }\textbf {\bibinfo {volume} {82}} (\bibinfo {year} {2010})}\BibitemShut
  {NoStop}%
\bibitem [{\citenamefont {Cai}\ \emph {et~al.}(2000)\citenamefont {Cai},
  \citenamefont {Painter},\ and\ \citenamefont {Vahala}}]{Cai2000}%
  \BibitemOpen
  \bibfield  {author} {\bibinfo {author} {\bibfnamefont {M.}~\bibnamefont
  {Cai}}, \bibinfo {author} {\bibfnamefont {O.}~\bibnamefont {Painter}}, \ and\
  \bibinfo {author} {\bibfnamefont {K.~J.}\ \bibnamefont {Vahala}},\
  }\href@noop {} {\bibfield  {journal} {\bibinfo  {journal} {Phys. Rev. Lett.}\
  }\textbf {\bibinfo {volume} {85}},\ \bibinfo {pages} {74} (\bibinfo {year}
  {2000})}\BibitemShut {NoStop}%
\bibitem [{\citenamefont {Faraon}\ \emph {et~al.}(2007)\citenamefont {Faraon},
  \citenamefont {Waks}, \citenamefont {Englund}, \citenamefont {Fushman},\ and\
  \citenamefont {Vu{\v{c}}kovi{\'c}}}]{Faraon2007}%
  \BibitemOpen
  \bibfield  {author} {\bibinfo {author} {\bibfnamefont {A.}~\bibnamefont
  {Faraon}}, \bibinfo {author} {\bibfnamefont {E.}~\bibnamefont {Waks}},
  \bibinfo {author} {\bibfnamefont {D.}~\bibnamefont {Englund}}, \bibinfo
  {author} {\bibfnamefont {I.}~\bibnamefont {Fushman}}, \ and\ \bibinfo
  {author} {\bibfnamefont {J.}~\bibnamefont {Vu{\v{c}}kovi{\'c}}},\ }\href@noop
  {} {\bibfield  {journal} {\bibinfo  {journal} {Appl. Phys. Lett.}\ }\textbf
  {\bibinfo {volume} {90}},\ \bibinfo {pages} {073102} (\bibinfo {year}
  {2007})}\BibitemShut {NoStop}%
\bibitem [{\citenamefont {Arcari}\ \emph {et~al.}(2014)\citenamefont {Arcari},
  \citenamefont {S{\"o}llner}, \citenamefont {Javadi}, \citenamefont {Hansen},
  \citenamefont {Mahmoodian}, \citenamefont {Liu}, \citenamefont {Thyrrestrup},
  \citenamefont {Lee}, \citenamefont {Song}, \citenamefont {Stobbe} \emph
  {et~al.}}]{Arcari2014}%
  \BibitemOpen
  \bibfield  {author} {\bibinfo {author} {\bibfnamefont {M.}~\bibnamefont
  {Arcari}}, \bibinfo {author} {\bibfnamefont {I.}~\bibnamefont {S{\"o}llner}},
  \bibinfo {author} {\bibfnamefont {A.}~\bibnamefont {Javadi}}, \bibinfo
  {author} {\bibfnamefont {S.~L.}\ \bibnamefont {Hansen}}, \bibinfo {author}
  {\bibfnamefont {S.}~\bibnamefont {Mahmoodian}}, \bibinfo {author}
  {\bibfnamefont {J.}~\bibnamefont {Liu}}, \bibinfo {author} {\bibfnamefont
  {H.}~\bibnamefont {Thyrrestrup}}, \bibinfo {author} {\bibfnamefont {E.~H.}\
  \bibnamefont {Lee}}, \bibinfo {author} {\bibfnamefont {J.~D.}\ \bibnamefont
  {Song}}, \bibinfo {author} {\bibfnamefont {S.}~\bibnamefont {Stobbe}},  \emph
  {et~al.},\ }\href@noop {} {\bibfield  {journal} {\bibinfo  {journal} {Phys.
  Rev. Lett.}\ }\textbf {\bibinfo {volume} {113}},\ \bibinfo {pages} {093603}
  (\bibinfo {year} {2014})}\BibitemShut {NoStop}%
\bibitem [{\citenamefont {Tiecke}\ \emph {et~al.}(2014)\citenamefont {Tiecke},
  \citenamefont {Thompson}, \citenamefont {de~Leon}, \citenamefont {Liu},
  \citenamefont {Vuleti{\'c}},\ and\ \citenamefont
  {Lukin}}]{tiecke2014nanophotonic}%
  \BibitemOpen
  \bibfield  {author} {\bibinfo {author} {\bibfnamefont {T.}~\bibnamefont
  {Tiecke}}, \bibinfo {author} {\bibfnamefont {J.~D.}\ \bibnamefont
  {Thompson}}, \bibinfo {author} {\bibfnamefont {N.~P.}\ \bibnamefont
  {de~Leon}}, \bibinfo {author} {\bibfnamefont {L.}~\bibnamefont {Liu}},
  \bibinfo {author} {\bibfnamefont {V.}~\bibnamefont {Vuleti{\'c}}}, \ and\
  \bibinfo {author} {\bibfnamefont {M.~D.}\ \bibnamefont {Lukin}},\ }\href@noop
  {} {\bibfield  {journal} {\bibinfo  {journal} {Nature}\ }\textbf {\bibinfo
  {volume} {508}},\ \bibinfo {pages} {241} (\bibinfo {year}
  {2014})}\BibitemShut {NoStop}%
\end{thebibliography}
%

\end{document}